\documentclass[prb,fleqn,12pt,showpacs]{revtex4-1}
\usepackage{epsfig}
\usepackage{graphicx}
\usepackage{amsfonts}
\begin{document} 

\title{Magnon self energy in the correlated ferromagnetic Kondo lattice model: 
spin-charge coupling effects on magnon excitations in manganites}

\author{Dheeraj Kumar Singh$^1$}
\author{Avinash Singh$^2$}
\email{avinas@iitk.ac.in} 
\affiliation{$^1$Asia Pacific Center for Theoretical Physics, Pohang, Gyeongbuk 790-784, Korea}
\affiliation{$^2$Department of Physics, Indian Institute of Technology Kanpur - 208016}
\begin{abstract}
Magnon self energy due to spin-charge coupling is calculated for the correlated ferromagnetic Kondo lattice model using a diagrammatic expansion scheme. Systematically incorporating correlation effects in the form of self-energy and vertex corrections, the expansion scheme explicitly preserves the continuous spin rotation symmetry and hence the Goldstone mode. Due to a near cancellation of the correlation-induced quantum correction terms at intermediate coupling and optimal band filling relevant for ferromagnetic manganites, the renormalized magnon energies for the correlated FKLM are nearly independent of correlation term.  Even at higher band fillings, despite exhibiting overall non-Heisenberg behavior, magnon dispersion in the $\Gamma$-X direction retains nearly Heisenberg form. Therefore, the experimentally observed doping dependent zone-boundary magnon softening must be ascribed to spin-orbital coupling effects. 
\end{abstract}

\maketitle
\newpage

\section{Introduction}
 
Magnetic excitations in the ferromagnetic Kondo lattice model (FKLM) have been investigated for various systems 
such as the colossal magnetoresistive manganites\cite{furukawa_96,wang_1998,vogt_2001,golosov_2000,shannon_2002,kapetanakis_2006,sudhakar0_2008}
heavy fermion materials,\cite{sigrist_92} ferromagnetic metals Gd and doped EuX,\cite{donath_98} 
ordered diluted ferromagnets,\cite{subrat_2005} and diluted magnetic semiconductors.\cite{asingh_2007,as_prb2007} 
Magnons in the FKLM yield important information about the emergent effective spin couplings generated by the exchange of the particle-hole propagator between the localized magnetic moments, the finite-temperature spin dynamics, and the zero-temperature  
magnon damping arising from the purely quantum spin-charge coupling effect in a band ferromagnet, \cite{sudhakar0_2008,sudhakar1_2008,dheeraj0_2010} which is totally absent in a localized spin model. 

Most investigations, however, often exclude the intra-orbital Coulomb interaction for the band fermion, which is especially 
inadequate for systems with highly correlated $d$-orbitals such as manganites. Indeed, photoemission and x-ray absorption experiments,\cite{Bocquet_1992,Saitoh_1995} as well as band structure calculations\cite{Satpathy_1996} yield $U\sim 3.5-8$ eV, whereas similar estimates for the Hund's coupling and bandwidth are $J\sim W\sim 2-3$ eV,\cite{Dagotto_2001} indicating the  importance of local correlations in the $e_g$ orbitals. 

In the few works where role of Coulomb interaction was investigated, one of the motivation was to account for the various anomalies in the magnon spectra observed in neutron scattering experiments on the ferromagnetic phase of colossal magnetoresistive manganites.\cite{hwang_98,dai_2000,tapan_2002,ye_2006,ye_2007,zhang_2007,moussa_2007} For instance, magnon dispersion shows significant zone-boundary softening in the $\Gamma$-X direction, indicating additional $-(1-\cos q_x)^2$ type non-Heisenberg term, underlining the limitation of the conventional double-exchange model. This feature is usually modeled by including a fourth neighbour interaction term $J_4 (1-\cos 2q_x)$ in the magnon dispersion, which contributes to the spin stiffness but not to the zone-boundary energy. Within a finite range of carrier density, the measured spin stiffness remains almost constant while the anomalous softening of the zone-boundary modes show substantial enhancement with increasing hole concentration.

Spin dynamics in the correlated FKLM has been investigated using the random phase approximation (RPA) with large Coulomb repulsion treated within Gutzwiller projection,\cite{Sun_2002} composite operator method,\cite{Mancini_2001} Holstein-Primakoff transformation,\cite{golosov_2005} and variational method.\cite{kapetanakis_2007} Gutzwiller projection in the mean-field approximation leads to modulation of hopping as function of carrier concentration, while magnon excitations are obtained in the RPA. In the self-consistent composite operator method, an extended FKLM including both intra-orbital Coulomb interaction as well as exchange interaction between localized spin has been explored, wherein both itinerant and localized contributions to magnetic couplings were treated equivalently. Features such as doping dependent asymmetry and sensitivity to Coulomb interaction of the stability of the ferromagnetic phase were emphasized.\cite{golosov_2005} Quantum corrections to magnon excitations beyond RPA were incorporated in the variational three-body calculation.

While hopping modulation within Gutzwiller projection rules out any zone-boundary softening, composite operator method yields enhanced softening in $\Gamma$-R as compared to $\Gamma$-X direction, in contrast with experiments.\cite{endoh_2005,ye_2007} Within the Holstein-Primakoff approach, overall non-Heisenberg feature ($\omega_X < \omega_M/2$) of magnon dispersion was linked to  anomalous zone-boundary softening, which was attributed entirely to the role of $U$, despite the absence of any magnon energy suppression in the $\Gamma$-X direction. On the other hand, while RPA magnon energy in the variational method shows nearly Heisenberg form, quantum corrections yield significant non-Heisenberg behavior, a feature also present in the uncorrelated FKLM.\cite{sudhakar0_2008,dheeraj_thesis} In the variational method, a strong doping dependence was obtained for the calculated spin stiffness near optimal doping, which is inconsistent with experiments. Both the Holstein-Primakoff and variational studies considered only the one-band model in two dimensions, and therefore focussed on the range 0.5$\le$$n$$\le$0.8 of band fillings $n$=1-$x$, corresponding to the metallic ferromagnetic regime (0.2$\le$$x$$\le$0.5). Including the realistic two-fold $e_g$ orbital degeneracy changes the optimal band filling (for $x$=0.3) to $n$=$(1-x)/2$=0.35 per orbital. 

The above contrasting results and limitations therefore necessitate a systematic investigation of the role of intra-orbital Coulomb repulsion on magnon excitations in the two-band correlated FKLM. In this paper, we will therefore investigate magnon self energy corrections in the correlated FKLM. The physically transparent purely fermionic representation of the FKLM will be employed which allows for a conventional many-body diagrammatic approach.\cite{sudhakar0_2008} Guided by the inverse-degeneracy-expansion scheme, correlation effects in the form of self-energy and vertex corrections are systematically incorporated so as to explicitly preserve the continuous spin rotation symmetry and hence the Goldstone mode.\cite{asingh} By treating both Hund's coupling ($J$) and the correlation term ($U$) on an equal footing, the diagrammatic approach is extended to the correlated FKLM, and quantum corrections to magnon excitations beyond the classical (RPA) level are obtained in two and three dimensions for different band fillings and interaction strengths.

\section{One band correlated FKLM}
We will consider the following purely fermionic representation for the correlated FKLM in terms of a two orbital Hubbard model involving a localized orbital ($\alpha$) for the core spin and a band orbital ($\beta$) with the correlation term included:  
\begin{eqnarray}
H &=&\sum_{i\sigma} \epsilon_\alpha a^\dagger _{i \alpha \sigma}  a_{i \alpha \sigma}
- U_\alpha \sum_i {\bf S}_{i \alpha}.{\bf S}_{i \alpha} \nonumber \\ 
&+& \sum_{{\bf k}\sigma} \epsilon_{\bf k} a^\dagger _{{\bf k}\beta \sigma}  a_{{\bf k}\beta \sigma}
-U\sum_{i} {\bf S}_{i \beta}.{\bf S}_{i \beta}
- 2J \sum_{i} {\bf S}_{i \alpha}.{\bf S}_{i \beta}
\end{eqnarray}
Here, the localized $\alpha$ orbital with large Hubbard interaction $U_{\alpha}$ yields exchange-split spin levels. With occupancies $n_{\alpha \uparrow}$=1 and $n_{\alpha \downarrow}$=0, the half-filled localized level with local magnetization $m_\alpha$=1 represents spin $S$=1/2 localized magnetic moments. Spin-$S$ magnetic moments can be similarly represented by including multiple ($N$=2$S$) $\alpha$ orbitals per site with Hund's coupling. The correlated $\beta$ band with dispersion $\epsilon_{\bf k}$=$-2t \sum_{i=1} ^d \cos k_i$ represent mobile fermions with Hubbard interaction $U$. The Hund's coupling $J$ represents the conventional exchange interaction between the localized spin and the mobile fermion spin, as considered phenomenologically in the ferromagnetic Kondo lattice model.

We consider a saturated ferromagnetic state, with ordering chosen in the $\hat{z}$ direction, and spatially uniform magnetizations $\langle S_{i\alpha}^z \rangle$=1/2 and $\langle  S_{i\beta}^z \rangle$=$m$/2 for the localized and mobile fermions, corresponding to fully occupied $\alpha\uparrow$ level and partially occupied $\beta\uparrow$ band. The interaction term at  Hartree-Fock level can be written as: 
\begin{equation}
H_{\rm int}^{\rm HF} = 
-\sum_{i\mu} \psi_{i\mu}^\dagger [{\bf \sigma}.{\bf \Delta_\mu}] \psi_{i\mu}
\end{equation}
where $\psi_\mu$ represents the fermion field operators for orbitals $\mu=\alpha,\beta$,
and the respective exchange splittings are:
\begin{eqnarray}
2\Delta_\alpha &=& 2(U_\alpha \langle S_{i\alpha}^z \rangle + J\langle  S_{i\beta}^z \rangle ) = U_\alpha + Jm \nonumber \\ 
2\Delta _\beta  &=& 2( {U\langle {S_{i\beta }^z } \rangle  + J\langle {S_{i\alpha }^z } \rangle }) = Um + J .
\end{eqnarray}

The bare antiparallel-spin particle-hole propagators for the two orbitals are given by:
\begin{eqnarray}
\chi^0 _\alpha (\omega) &=& 
\frac{1}{2\Delta_\alpha +\omega -i \eta} = 
\frac{1}{U_\alpha + Jm +\omega -i \eta} \nonumber \\
\chi^0 _\beta ({\bf q},\omega) &=& \sum_{\bf k}
\frac{1}{\epsilon_{\bf k-q}^{\downarrow +} - \epsilon_{\bf k}^{\uparrow -} + \omega -i \eta} 
= \frac{m}{Um+J + \omega -i \eta} \;\;\; ({\rm for} \;\; {\bf q}=0) \; ,
\end{eqnarray}
where $\epsilon_{\bf k}^\sigma = \epsilon_{\bf k} - \sigma \Delta_\beta$ 
are the exchange-split band energies, and the superscripts $+$ $(-)$ refer to particle (hole) states above (below) the Fermi energy $\epsilon_{\rm F}$.

\begin{figure}
\begin{center}
\vspace*{-2mm}
\hspace*{0mm}
\psfig{figure=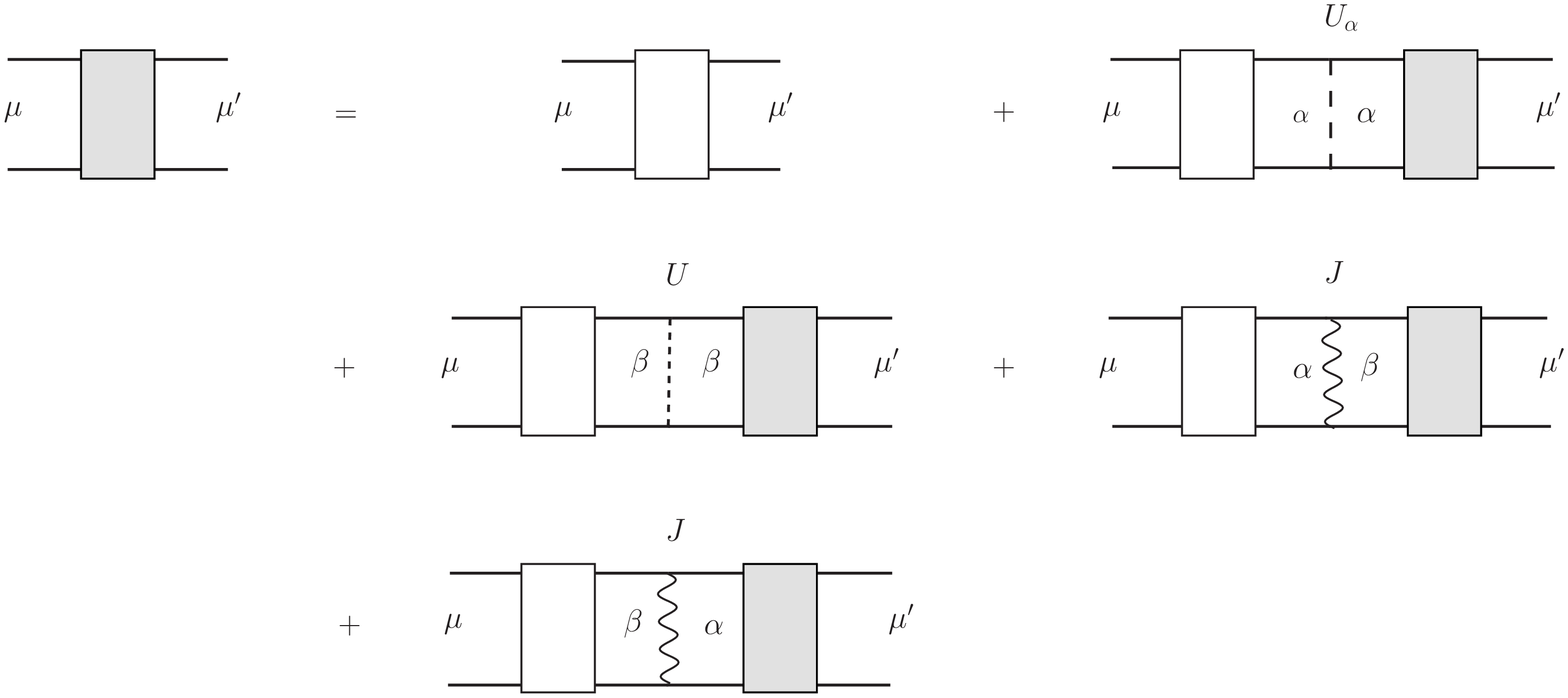,width=150mm,angle=0}
\vspace*{-5mm}
\end{center}
\caption{Exact diagrammatic representation of the coupled equations for the components of $\chi^{-+}$ (shaded box) in terms of the irreducible particle-hole propagator $\phi$ (open box).}
\label{prop}
\end{figure}

\begin{figure}
  \begin{center}
    \begin{tabular}{cc}
      \resizebox{100mm}{!}{\includegraphics[angle=0]{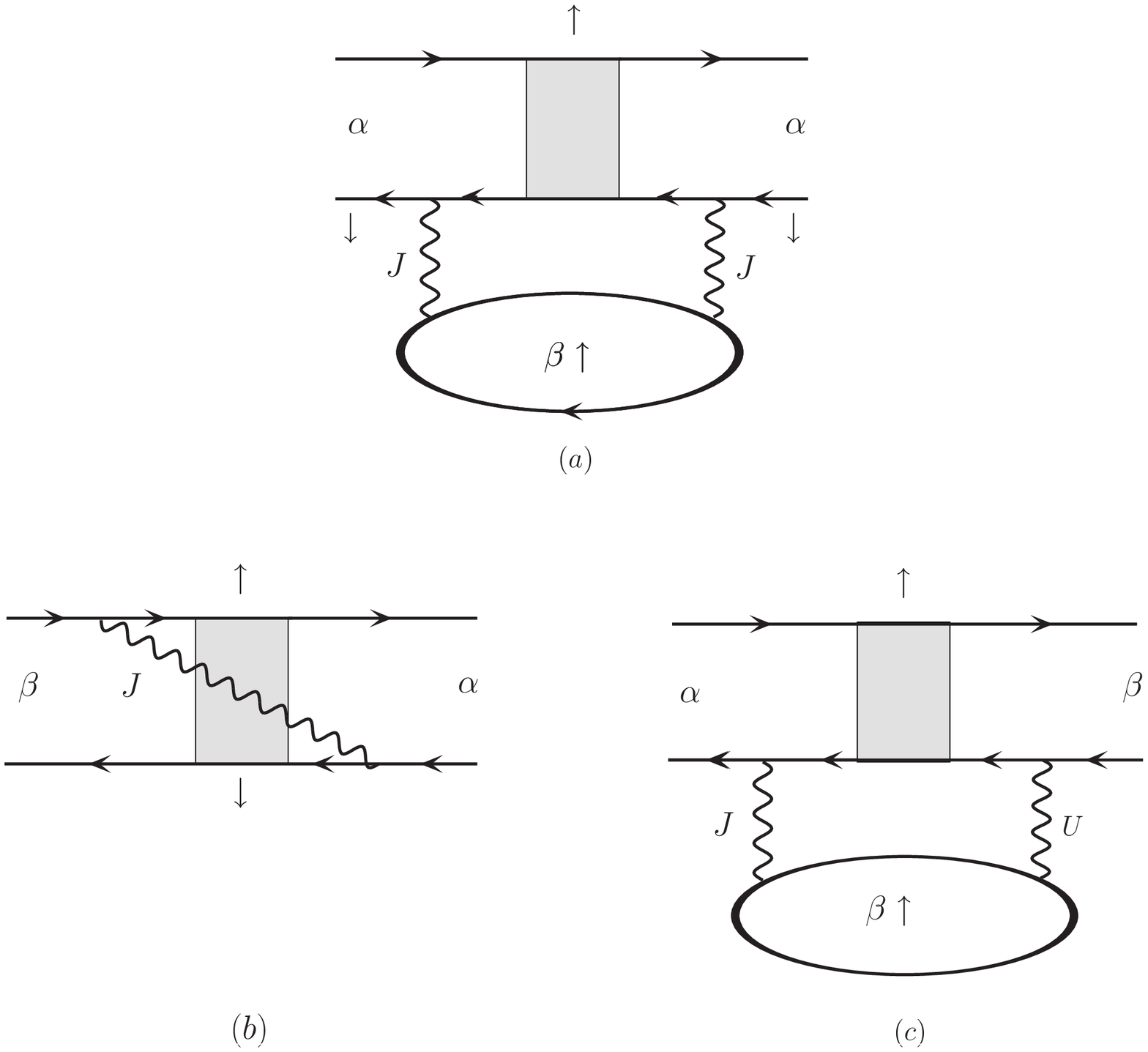}} \\
      \resizebox{100mm}{!}{\includegraphics[angle=0]{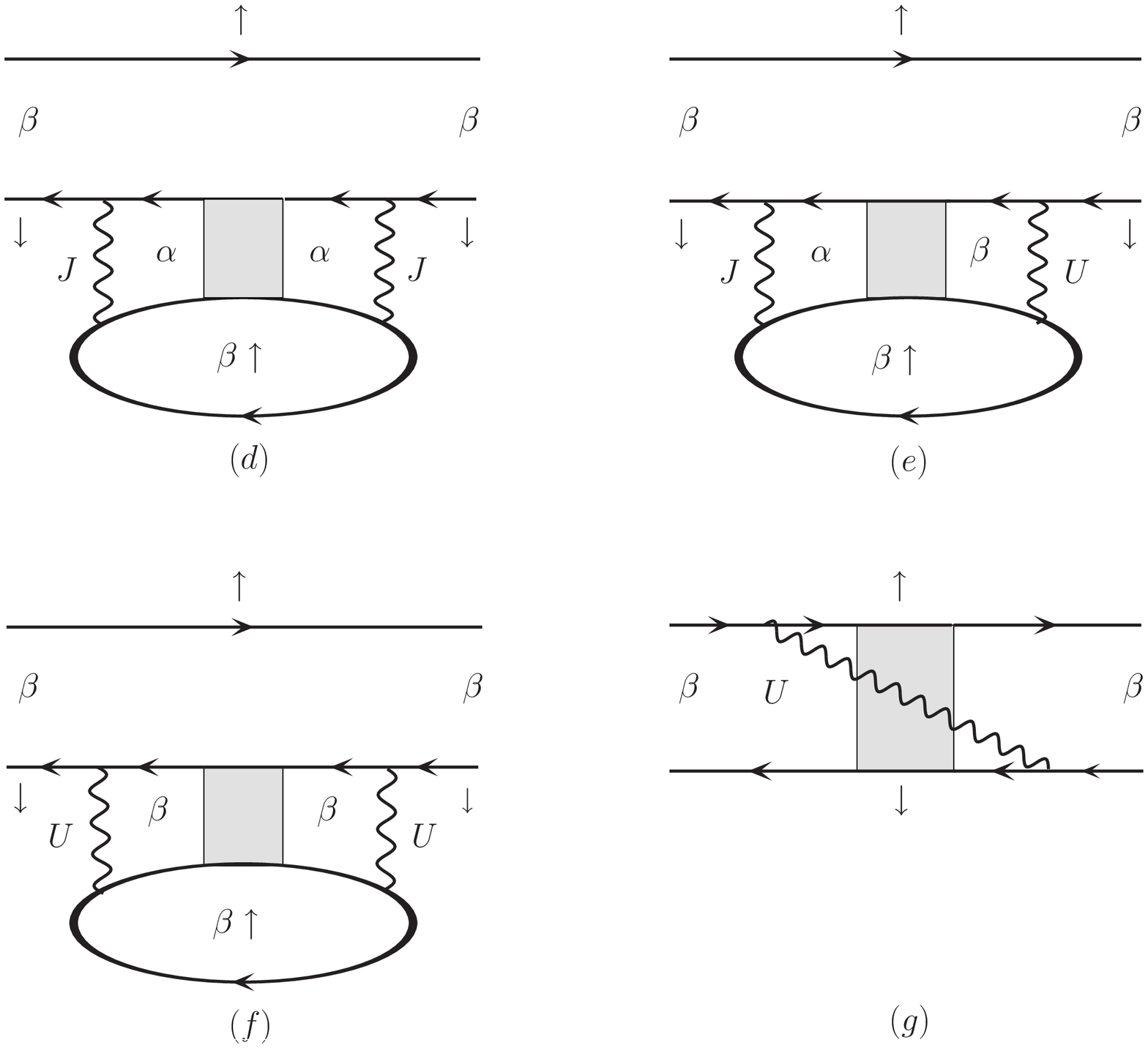}} \\
      \resizebox{100mm}{!}{\includegraphics[angle=0]{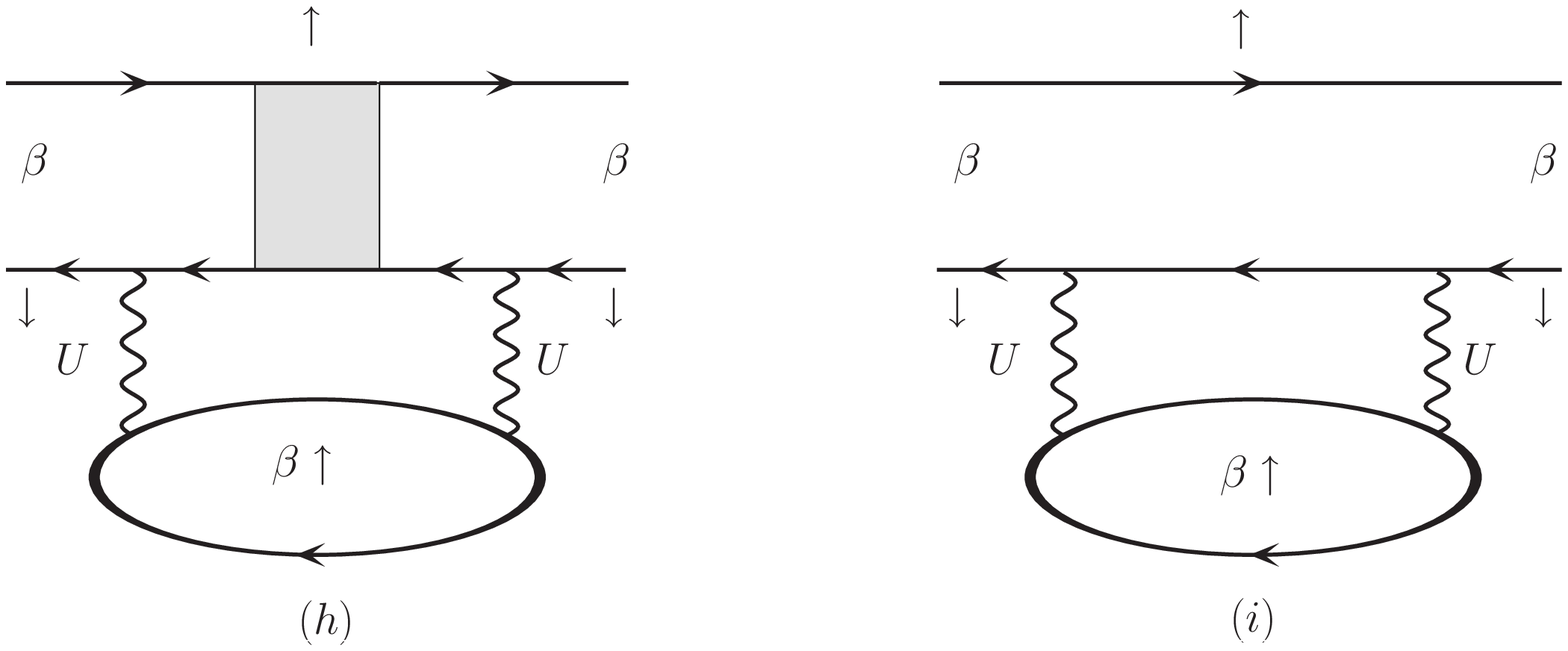}} \\
    \end{tabular}
    \caption{First order quantum corrections to the irreducible particle-hole propagator $\phi$.}
   \label{phi1}
  \end{center}
\end{figure}

\subsection{Magnon propagator}
The different components $(\mu,\mu' = \alpha,\beta)$ 
of the time-ordered magnon propagator in the ferromagnetic ground state $| \Psi_0 \rangle$ are given by:  
\begin{equation}
\chi ^{-+} _{\mu\mu'} ({\bf q},\omega) =
i \int dt  \; e^{i\omega (t-t')} \sum_j e^{i{\bf q}.({\bf r}_i - {\bf r}_j)} 
\langle \Psi_0 |{\rm T}[S_{i\mu} ^- (t) S_{j\mu'} ^+ (t')]\Psi_0 \rangle 
\end{equation}
in terms of spin-lowering and spin-raising operators $S_{i\mu} ^\mp = \psi_{i\mu} ^\dagger (\sigma^\mp/2) \psi_{i\mu}$.
As shown diagrammatically in Fig. \ref{prop}, the coupled equations for different components of $\chi ^{-+} $ can be expressed exactly in terms of the irreducible particle-hole propagators $\phi_{\mu \mu^\prime}$: 
\begin{eqnarray}
\chi ^{-+} _{\mu\mu'} &=& \phi_{\mu\mu'} + \phi_{\mu\alpha} U_\alpha \chi ^{-+} _{\alpha\mu'} + \phi_{\mu\beta} U \chi ^{-+} _{\beta\mu'} + \phi_{\mu\alpha} J \chi ^{-+} _{\beta\mu'} + \phi_{\mu\beta} J \chi ^{-+} _{\alpha\mu'}. 
\end{eqnarray}

Guided by the inverse-degeneracy $(1/\cal N)$ expansion scheme for the $\cal N$-orbital Hubbard model, the irreducible propagator can be systematically expanded as:  
\begin{eqnarray}
\phi_{\mu\mu'} &=& \chi^0 _\mu \delta_{\mu\mu'} + \phi^{(1)}_{\mu\mu'} + \phi^{(2)}_{\mu\mu'} + ...
\end{eqnarray}
where $\chi^0$ is the bare particle-hole propagator, and correlation effects in the form of self-energy and vertex corrections are incorporated systematically in the quantum corrections $\phi^{(1)}$, $\phi^{(2)}$ etc., so that spin-rotation symmetry and hence the Goldstone mode are explicitly preserved order by order. 

Solving the coupled Eq. (6) for different components, we obtain:
\begin{eqnarray}
\chi^{-+}_{\beta\beta} &=& \frac {\phi_\beta(1-U_\alpha \phi_\alpha) +U_\alpha \phi_{\alpha \beta} \phi_{\beta \alpha}}{(1-U_\alpha \phi_\alpha) (1-U \phi_\beta)-J^2\phi_\alpha \phi_\beta -J(\phi_{\alpha\beta}+\phi_{\beta\alpha})+(J^2-U_\alpha U)\phi_{\alpha\beta}\phi_{\beta \alpha}} 
\end{eqnarray}
\begin{eqnarray}
\chi^{-+}_{\alpha\beta} &=& \frac {\phi_{\alpha \beta}+J\phi_{\alpha}\phi_{\beta}-J\phi_{\alpha \beta}\phi_{\beta \alpha}}
{(1-U_\alpha \phi_\alpha)(1-U \phi_\beta)-J^2\phi_\alpha \phi_\beta -J(\phi_{\alpha\beta}+\phi_{\beta\alpha})+(J^2-U_\alpha U)\phi_{\alpha\beta} \phi_{\beta \alpha}} 
\end{eqnarray}
\begin{eqnarray}
\chi^{-+}_{\alpha\alpha} &=& \frac {\phi_\alpha(1-U \phi_\beta)+U \phi_{\alpha \beta} \phi_{\beta \alpha}}{(1-U_\alpha \phi_\alpha)
(1-U \phi_\beta)-J^2\phi_\alpha \phi_\beta-J(\phi_{\alpha\beta}+\phi_{\beta\alpha})+(J^2-U_\alpha U )\phi_{\alpha\beta}\phi_{\beta \alpha}}
 \nonumber \\
&\approx& \frac {\phi_\alpha(1-U \phi_\beta)}{(1-U_\alpha \phi_\alpha)(1-U \phi_\beta)-J^2\phi_\alpha \phi_\beta-J(\phi_{\alpha\beta} +\phi_{\beta\alpha})+(J^2-U_\alpha U)\phi_{\alpha\beta}\phi_{\beta \alpha}} \nonumber \\
&=& \frac {1}{(g_\alpha)^{-1}-J^2g_\beta-2J\phi_{\alpha \beta}[\phi_\alpha(1-U \phi_\beta)]^{-1}+(J^2-U_\alpha U)\phi_{\alpha\beta}^2 [\phi_\alpha(1-U\phi_\beta)]^{-1}}
\end{eqnarray}
where 
\begin{equation}
g_\alpha (\omega) \equiv \frac{\phi_{\alpha \alpha}}{1-U_\alpha \phi_{\alpha \alpha}} \ {\rm and} \ g_\beta (\omega) \equiv \frac{\phi_{\beta \beta}}{1-U \phi_{\beta \beta}}
\end{equation} 
represent the magnon propagators for the localized and mobile electrons. Resulting from the coupled nature of Eq. (6) for the different components, the common denominator of different components of the propagator ensures a single Goldstone mode for all, as expected. With only the lowest-order term $\phi = \chi^0_{\mu \mu^\prime}$, Eqs. (8)-(10) yield the different components in the ladder-sum (RPA) approximation, and are given in the Appendix. 

\subsection{Quantum corrections and magnon self-energy}

The first-order quantum corrections to the different components of the irreducible propagator are shown diagrammatically in Fig. \ref{phi1}. Both self energy and vertex corrections are included. The shaded blocks correspond to the RPA-level vertices $[\Gamma^{-+}_{\mu\nu}]$ and propagators $[\chi^{-+}_{\mu\nu}]$ (Appendix). The corresponding expressions are obtained as:
\begin{eqnarray}
\phi^{(a)}_{\alpha\alpha}({\bf q},\omega) &=& J^2\left(\frac{1}{2\Delta_\alpha+\omega}\right)^2 \sum_{\bf Q} \int \frac {d\Omega}{2\pi i}\; [\chi^{-+}_{\alpha\alpha}]_{\rm RPA}({\bf Q},\Omega)\nonumber\\
& &\times \sum_{\bf k}\left(\frac{1}{\epsilon^{\uparrow +}_{\bf k-q+Q}-\epsilon^{\uparrow -}_{\bf k}+\omega-\Omega-i\eta}\right)\nonumber\\
\phi^{(b)}_{\beta\alpha} ({\bf q},\omega) &=& -J \left(\frac{1}{2\Delta_\alpha+\omega}\right) \sum_{\bf Q}
\int \frac {d\Omega}{2\pi i}\; \Gamma^{-+}_{\beta\alpha}({\bf Q},\Omega) \chi^{\circ}_\alpha (\Omega) \nonumber\\
& &\times \sum_{\bf k}
\left(\frac{1}{\epsilon^{\downarrow +}_{\bf k-q}-\epsilon^{\uparrow -}_{\bf k}+\omega -i\eta}\right)
\left(\frac{1}{\epsilon^{\uparrow +}_{\bf k-q+Q}-\epsilon^{\uparrow -}_{\bf k}+\omega-\Omega-i\eta}\right)\nonumber\\
\phi^{(c)}_{\beta\alpha} ({\bf q},\omega) &=& JU \left(\frac{1}{2\Delta_\alpha+\omega}\right) \sum_{\bf Q} 
\int \frac {d\Omega}{2\pi i}\; \Gamma^{-+}_{\beta\alpha}({\bf Q},\Omega) \chi^{\circ}_\alpha (\Omega) \nonumber\\
& &\times \sum_{\bf k'} 
\left(\frac{1}{\epsilon^{\downarrow +}_{\bf k'-q}-\epsilon^{\uparrow -}_{\bf k'}+\omega -i\eta}\right)
\left(\frac{1}{\epsilon^{\downarrow +}_{\bf k'-Q}-\epsilon^{\uparrow -}_{\bf k'}+\Omega -i\eta}\right) \nonumber \\ 
& &\times \sum_{\bf k}
\left(\frac{1}{\epsilon^{\uparrow +}_{\bf k-q+Q}-\epsilon^{\uparrow -}_{\bf k}+\omega-\Omega-i\eta}\right)\nonumber\\ 
\phi^{(d)}_{\beta\beta}+\phi^{(e)}_{\beta\beta}+\phi^{(f)}_{\beta\beta}({\bf q},\omega) &=& 
\sum_{\bf Q} \int \frac{d\Omega}{2\pi i} \nonumber\\
& & \times \left\{J^2[\chi^{-+}_{\alpha\alpha}]_{\rm RPA}({\bf Q}, \Omega)+2JU[\chi^{-+}_{\alpha\beta}]_{\rm RPA}({\bf Q}, \Omega)+U^2 [\chi^{-+}_{\beta\beta}]_{\rm RPA}({\bf Q}, \Omega)\right\} \nonumber \\
 & & \times \sum_{\bf k}
\left(\frac{1}{\epsilon^{\downarrow +}_{\bf k-q}-\epsilon^{\uparrow -}_{\bf k}+\omega -i\eta}\right)^2
\left(\frac{1}{\epsilon^{\uparrow +}_{\bf k-q+Q}-\epsilon^{\uparrow -}_{\bf k}+\omega-\Omega-i\eta}\right)\nonumber\\
\phi^{(g)}_{\beta\beta} ({\bf q},\omega) &=& -2U \sum_{\bf Q} \int \frac {d\Omega}{2\pi i}\; \Gamma^{-+}_{\beta\beta}({\bf Q},\Omega) \nonumber \\
& &\times \sum_{\bf k} \left(\frac{1}{\epsilon^{\downarrow +}_{\bf k-q}-\epsilon^{\uparrow -}_{\bf k}+\omega -i\eta}\right)  \left(\frac{1}{\epsilon^{\uparrow +}_{\bf k-q+Q}-\epsilon^{\uparrow -}_{\bf k}+\omega-\Omega-i\eta}\right)\nonumber \\
& &\times \sum_{\bf k'} \left(\frac{1}{\epsilon^{\downarrow +}_{\bf k'-q}-\epsilon^{\uparrow -}_{\bf k'}+\omega -i\eta}\right)
\left(\frac{1}{\epsilon^{\downarrow +}_{\bf k'-Q}-\epsilon^{\uparrow -}_{\bf k'}+\Omega -i\eta}\right)\nonumber\\
\phi^{(h)}_{\beta\beta}({\bf q},\omega) &=& U^2\sum_{\bf Q} \int \frac {d\Omega}{2\pi i}\; \Gamma^{-+}_{\beta\beta}({\bf Q},\Omega)\nonumber \\
& &\times \left[\sum_{\bf k'} \left(\frac{1}{\epsilon^{\downarrow +}_{\bf k'-q}-\epsilon^{\uparrow -}_{\bf k'}+\omega -i\eta}\right)\left(\frac{1}{\epsilon^{\downarrow +}_{\bf k'-Q}-\epsilon^{\uparrow -}_{\bf k'}+\Omega -i\eta}\right)\right]^2 \nonumber\\
& &\times \sum_{\bf k} \left(\frac{1}{\epsilon^{\uparrow +}_{\bf k-q+Q}-\epsilon^{\uparrow -}_{\bf k}+\omega-\Omega-i\eta}\right)\nonumber \\
\phi^{(i)}_{\beta\beta}({\bf q},\omega) &=& U^2\sum_{\bf Q}\int \frac {d\Omega}{2\pi i} \sum_{\bf k'}\left(\frac{1}{\epsilon^{\downarrow +}_{\bf k'-q}-\epsilon^{\uparrow -}_{\bf k'}+\omega -i\eta}\right)^2 \nonumber \\ 
& & \times \left(\frac{1}{\epsilon^{\downarrow +}_{\bf k'-Q}-\epsilon^{\uparrow -}_{\bf k'}+\Omega -i\eta}\right) \sum_{\bf k}
\left(\frac{1}{\epsilon^{\uparrow +}_{\bf k-q+Q}-\epsilon^{\uparrow -}_{\bf k}+\omega-\Omega-i\eta}\right)\; . \nonumber \\
\end{eqnarray}

Separating out the bare part from the quantum corrections in the irreducible particle-hole propagators $\phi$ in (10), 
and dropping explicitly second-order terms such as $\phi^{(1)}_{\alpha\beta} \phi^{(1)}_{\beta\alpha}$, we obtain for the $\alpha\alpha$ component: 
\begin{eqnarray} 
\chi^{-+}_{\alpha\alpha}({\bf q},\omega) = \frac{1}
{[\omega + Jm - J^2 \chi^0 _\beta ({\bf q},\omega)/(1-U\chi^0 _\beta ({\bf q},\omega))] - 
[\Sigma_\alpha ({\bf q},\omega) + \Sigma_\beta ({\bf q},\omega)
+ 2\Sigma_{\alpha\beta} ({\bf q},\omega)]}. \nonumber \\
\end{eqnarray}
The zeroth-order first term in the denominator yields the RPA result. The Goldstone mode at this level is easily verified, confirming the systematic nature of the expansion. For $q = 0$, the bare fermion propagator $\chi^0 _\beta(0,\omega) = m/(Um + J + \omega)$, which yields a pole at $\omega =0$, as expected from the continuous spin-rotation symmetry.
For finite ${\bf q}$, the RPA magnon energy is obtained by solving the pole equation
\begin{equation}
\omega+Jm -J^2 \frac{ \chi^0 _\beta({\bf q},\omega)}{1-U\chi^0 _\beta ({\bf q},\omega)} = 0 .
\end{equation}
As $J^2 \chi^0 _\beta (0,0)/({1-U\chi^0 _\beta (0,0)}) = Jm$ from (4), and the $\omega$ dependence of the bare fermion propagator $\chi^0_\beta$ is relatively weak, the RPA magnon energy is approximately obtained as: 
\begin{equation}
\omega_{\bf q} ^0 = J^2 (2S) \left[ \frac{\chi^0 _\beta(0)}{1-U\chi^0 _\beta(0)} - \frac{\chi^0 _\beta({\bf q})}
{1-U\chi^0 _\beta({\bf q})} \right] 
\end{equation}
for the spin-$S$ case. For $J\sim W$, the magnon dispersion is nearly identical as for the uncorrelated FKLM, as shown in the Appendix. For smaller $J$ values, when the exchange gap $2JS+Um$ is comparable to bandwidth $W$, quantitative analysis does show overall non-Heisenberg behavior ($\omega_X < \omega_R /3$, i.e. magnon softening at X relative to R), arising from 2nd and 3rd neighbor effective spin couplings. However, strictly in the $\Gamma$-X direction, the dispersion remains nearly of the Heisenberg form $(1- \cos q_x)$. 
 
\begin{figure}
\begin{center}
\vspace*{-2mm}
\hspace*{0mm}
\psfig{figure=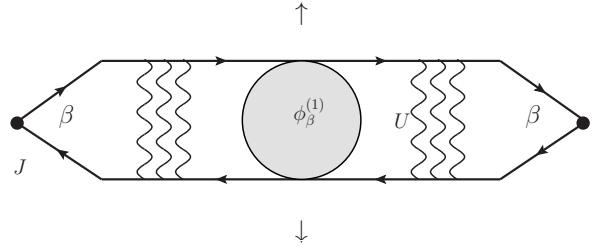,width=80mm,angle=0}
\vspace*{-5mm}
\end{center}
\caption{Effective vertex renormalization of the bare spin-fermion coupling due to particle-hole ladders in the correlated $\beta$ band.}
\label{vrenorm}
\end{figure}

Returning to Eq. (13), the first-order magnon self energies are obtained as: 
\begin{eqnarray}
\Sigma^{(1)}_\alpha ({\bf q},\omega) &=& (2 \Delta_\alpha +\omega)^2 \phi_{\alpha\alpha} ^{(1)}({\bf q},\omega) \nonumber \\
\Sigma^{(1)}_\beta  ({\bf q},\omega) &=& J'^2({\bf q},\omega) \phi_{\beta\beta} ^{(1)}({\bf q},\omega) \nonumber \\
\Sigma^{(1)}_{\alpha\beta} ({\bf q},\omega) &=& J'({\bf q},\omega)(2\Delta_\alpha+\omega) \phi_{\beta\alpha} ^{(1)}({\bf q},\omega).
\end{eqnarray}
These expressions are structurally similar as for the uncorrelated FKLM,\cite{sudhakar0_2008} 
with $J$ simply replaced by: 
\begin{equation} 
J'({\bf q},\omega)=\frac{J}{1-U\chi^0_\beta({\bf q},\omega)} 
\end{equation}
which represents an effective vertex renormalization of the spin-fermion coupling due to the particle-hole ladder sum in the correlated $\beta$-band as shown in Fig. \ref{vrenorm}. 

\subsection{Spin-charge coupling and magnon energy renormalization}

Using the relations between different components of RPA-level propagators and vertices (Appendix), the nine diagrams in Fig. \ref{phi1} can be combined into the following compact spin-charge coupling structure for the total magnon self-energy (Eq. 13):  
\begin{eqnarray}
\Sigma_{\rm magnon}({\bf q},\omega) &=& 
\Sigma_\alpha({\bf q},\omega) + \Sigma_\beta({\bf q},\omega) + \Sigma_{\alpha \beta}({\bf q},\omega) \nonumber \\
&=& \sum_{\bf k,Q} \int \frac{d\Omega}{2\pi i} \;
[\chi^{-+}_{\alpha\alpha}]_{\rm RPA} ({\bf Q},\Omega) \; \Gamma^2 \; \Pi^0 ({\bf k},{\bf q-Q},\omega-\Omega)
\end{eqnarray}
where $[\chi^{-+}_{\alpha\alpha}]_{\rm RPA}$ is given by Eq. (31) in the Appendix, $\Gamma$ represents the spin-charge coupling interaction vertex, and $\Pi^0$ is the charge fluctuation propagator given by: 
\begin{equation}
\Pi^0  ({\bf k};{\bf q-Q},\omega-\Omega) = \frac{1}
{\epsilon_{\bf k-q+Q}^{\uparrow +} - \epsilon_{\bf k}^{\uparrow -} + \omega - \Omega - i \eta }. 
\end{equation}

This correlation-induced coupling between spin and charge fluctuations arises from the scattering of magnon (energy $\omega^0_{\bf q}$) into intermediate spin-excitation states accompanied by charge fluctuations in the majority spin band. The intermediate states include both magnons (energy $\omega^0_{\bf Q}$) and Stoner excitations. The spin-charge coupling is similar to three-body correlations (between Fermi sea electron-hole pair and magnon) considered in variational investigation.\cite{kapetanakis_2007}

The spin-charge coupling interaction vertex is given by: 
\begin{eqnarray}
\Gamma({\bf k};{\bf q},\omega;{\bf Q},\Omega)&=& J \left(1-\frac {2J'({\bf q},\omega)S}{\epsilon_{\bf k-q}^{\downarrow +}
-\epsilon_{\bf k}^{\uparrow -} +\omega -i \eta} \; \mathit{f}_U ({\bf k};{\bf q},\omega;{\bf Q},\Omega)\right) 
\end{eqnarray}
 
\begin{eqnarray}
\mathit{f}_{U}({\bf k};{\bf q},\omega;{\bf Q},\Omega)&=&
\left(\frac {1-U\sum_{\bf k'}\chi^0_\beta({\bf k'};{\bf q},\omega)\chi^0_\beta({\bf k'};{\bf Q},\Omega)/\chi^0_\beta
({\bf k};{\bf q},\omega)}{1-U \chi^0_\beta({\bf Q},\Omega)}\right) \; .
\end{eqnarray}
The factor $f_U$ represents a correlation-induced correction, with
\begin{equation}
\chi^0_\beta({\bf k};{\bf q},\omega)= \frac{1}
{\epsilon_{\bf k-q}^{\downarrow +} - \epsilon_{\bf k}^{\uparrow -} + \omega - i\eta} = \frac{1}{2\Delta_\beta+\omega}
 \;\;\; ({\rm for} \;\; {\bf q}=0) \; .
\end{equation} 
The generalization for the case of localized spin-$S$ magnetic moments has been made by considering $2S$ localized orbitals and replacing the fermion-magnon interaction vertex $J$ by $J\sqrt{2S}$, so that the exchange gap is now given by $2\Delta_\beta$=$2JS$+$mU$.

The interaction vertex $\Gamma$ has three terms for the correlated FKLM, having nominal coefficients 1, $J$, and $JU$. The correspondence of resulting nine terms in magnon self energy (Eq. 18) to the nine quantum correction diagrams in Fig. 2 is as follows. The $J^2 U^2$, $2J^2 U$, and $J^2$ terms correspond to diagrams (h), (g), and (d)+(e)+(f), respectively. Similarly, the $2JU$ and $2J$ terms correspond to diagrams (c) and (b), respectively. Finally, the 1 term corresponds to diagram (a).

Two limiting cases for the interaction vertex are of interest. For $U$=0, $\mathit{f}_U$=1, and Eq. (20) reduces to:
\begin{equation}
\Gamma({\bf k};{\bf q},\omega;{\bf Q},\Omega) = J \left(1-\frac {2JS}{\epsilon_{\bf k-q}^{\downarrow +}
-\epsilon_{\bf k}^{\uparrow -} +\omega -i \eta}\right)
\end{equation}
as for the uncorrelated FKLM. For finite $U$, and $q,\omega$=0, since Goldstone mode condition is already exhausted at RPA level, magnon self energy must identically vanish. From Eqs. (21,22) we obtain:  
\begin{equation}
\mathit{f}_{U}({\bf k};0,0;{\bf Q},\Omega) = \left(\frac {1-U \sum_{\bf k'}\chi^0_\beta({\bf k'};{\bf Q},\Omega)}{1-U \chi^0_\beta({\bf Q},\Omega)}\right) = 1 . 
\end{equation} 
Furthermore, from Eqs. (17) and (20): 
\begin{equation} 
J'(0, 0) = \frac{J}{1-U \chi^0_\beta(0, 0)} = \frac{2J\Delta_\beta}{2JS} 
\end{equation}
and 
\begin{equation}
\Gamma({\bf k};0,0;{\bf Q},\Omega) = J \left(1-\frac {2J'(0,0)S}{2\Delta_\beta}\right) = 0 . 
\end{equation}
Thus, the interaction vertex and magnon self energy identically vanish, thereby ensuring the Goldstone mode.


\subsection{Is the correlation term significant?}

The correlation term renormalizes the spin-charge coupling magnon self energy in two distinct ways: (i) replacement of bare spin-fermion interaction by $J'({\bf q},\omega) = J/(1-U\chi^0 _\beta ({\bf q},\omega))$, and (ii) correlation factor $f_U$ which accounts for finite-$U$ induced self energy and vertex corrections in Fig. 2. So how does this affect the magnon self energy corrections and therefore the renormalized magnon energies? We consider here the physically relevant intermediate coupling 
($J$$\sim$$W$) and low band filling ($m$=$n$$\ll$1) case, as appropriate for the optimally doped manganites with $n$=0.35.

Distribution of the particle-hole term $\chi^0 _\beta ({\bf k};{\bf q},\omega)$ over allowed momentum range ($\epsilon_{\bf k} ^\uparrow < E_{\rm F}$) shows sharply peaked behavior, indicating nearly constant value which follows from nearly momentum independent denominator in Eq. (22), as the exchange gap $2JS+Um \gg W$, the electronic bandwidth. Therefore, in ${\bf k}$ summations, this term can be approximately replaced by a constant, the nearly flat  momentum-space behavior implying nearly local particle-hole fluctuations. Two of these terms in Eq. (21) thus get cancelled out, leading to:
\begin{equation}
\mathit{f}_{U}({\bf k};{\bf q},\omega;{\bf Q},\Omega) \approx \left(\frac {1-U \sum_{\bf k'}\chi^0_\beta({\bf k'};{\bf Q},\Omega)}{1-U \chi^0_\beta({\bf Q},\Omega)}\right) = 1.
\end{equation} 
Moreover, within same approximation:
\begin{equation}
\chi^0 _\beta ({\bf q},\omega) = \sum_{\bf k} \chi^0_\beta({\bf k};{\bf q},\omega)  
\approx m\chi^0_\beta({\bf k};{\bf q},\omega)
\end{equation}
as the normalized ${\bf k}$ summation over occupied states simply yields a factor $n=m$. 

Consequently, the expression for the spin-charge coupling interaction vertex (Eq. 20) approximately reduces to:
\begin{eqnarray}
\Gamma({\bf k};{\bf q},\omega;{\bf Q},\Omega) &=& J \left(1-\frac {2JS \chi^0 _\beta ({\bf k};{\bf q},\omega)}{1-U \chi^0 _\beta ({\bf q},\omega)}\mathit{f}_U ({\bf k};{\bf q},\omega;{\bf Q},\Omega)\right) 
\;\;\;\; ({\rm Eqs.} \;17\; {\rm and}\; 22) \nonumber \\ 
& \approx & J \left(1-\frac {2JS}{[\chi^0 _\beta ({\bf k};{\bf q},\omega)]^{-1} - Um }\right) \;\;\;\; ({\rm Eq.} \;27 \; {\rm and} \; 28) \nonumber \\
&=& J \left(1-\frac {2JS}{\epsilon_{\bf k-q}^{\downarrow +} - \epsilon_{\bf k}^{\uparrow -} +\omega -i \eta}\right) \;\;\;\; ({\rm Eq.} \;22) \; . 
\end{eqnarray}

In the last step of Eq. (29), due to cancellation of the $Um$ term, the effective exchange gap is 2JS as for the uncorrelated FKLM. Therefore, even for finite $U$, the spin-charge coupling vertex expression approximately reduces to that for the uncorrelated FKLM. In other words, quantum corrections to magnon excitations and magnon damping in the correlated FKLM is quantitatively similar as for the uncorrelated FKLM. This is in sharp contrast with earlier investigations which attributed zone-boundary softening entirely to the role of Coulomb interaction.\cite{golosov_2005,kapetanakis_2007} 

To summarize, in the physically relevant regime, while the ladder-sum renormalization of the spin-fermion vertex is exactly cancelled by the exchange gap renormalization, the approximately local nature of particle-hole fluctuations in the correlation factor leaves the spin-charge coupling unaffected. 

\begin{figure}
\begin{center}
\vspace*{-2mm}
\hspace*{0mm}
\psfig{figure=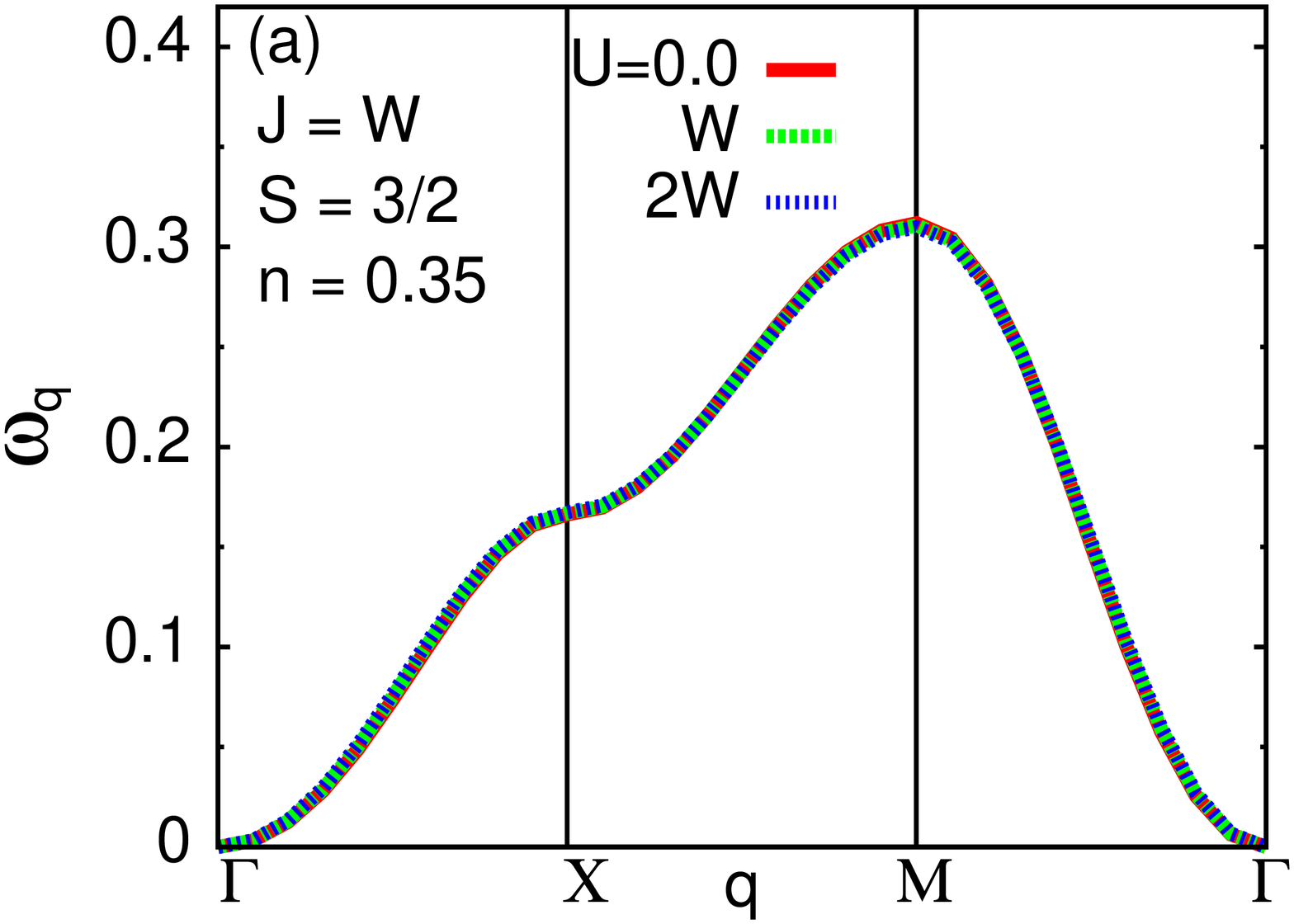,width=37.5mm,angle=-90}
\psfig{figure=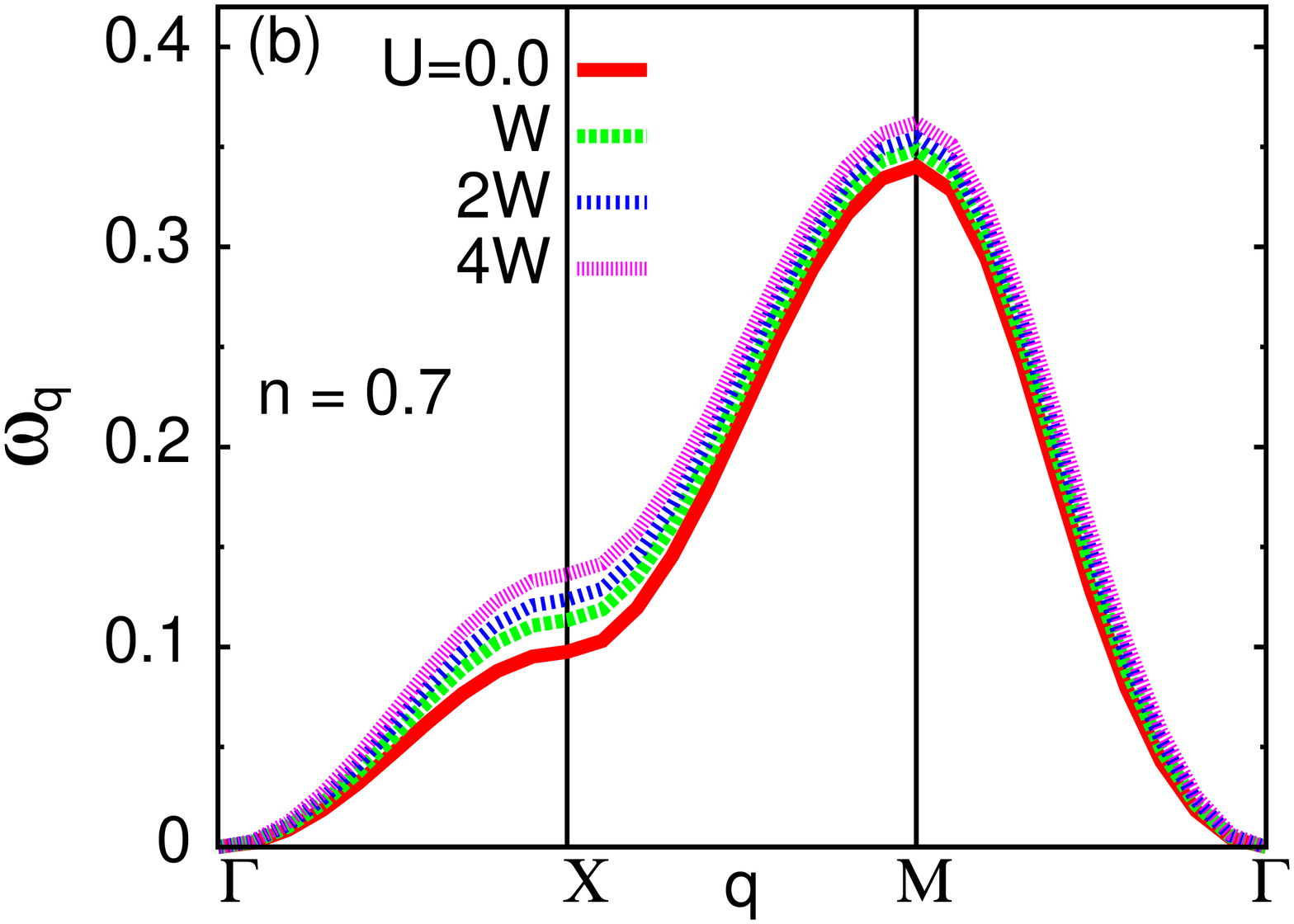,width=37.5mm,angle=-90}
\psfig{figure=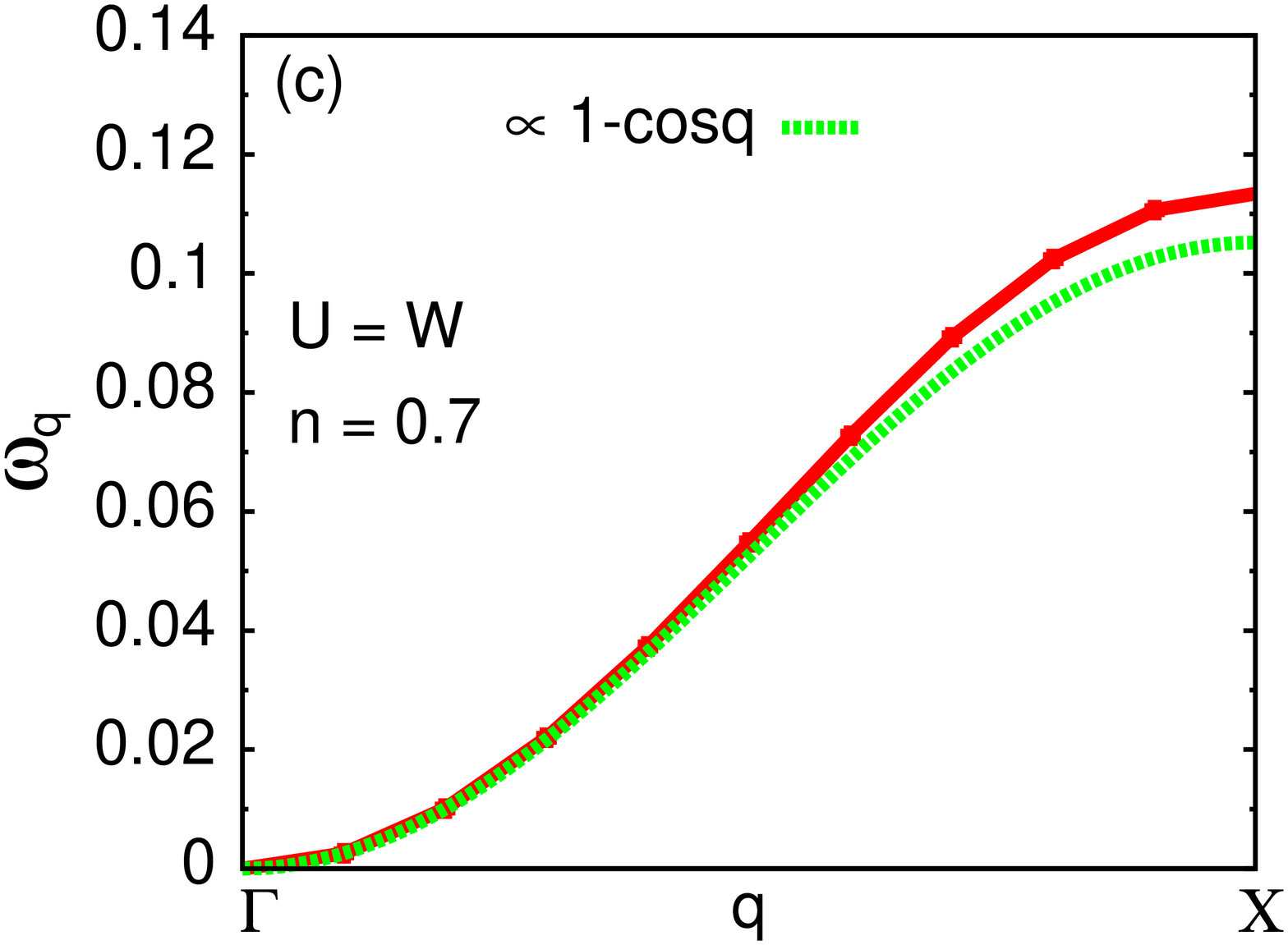,width=37.5mm,angle=-90}
\vspace*{-5mm}
\end{center}
\caption{Renormalized magnon energy in the one orbital FKLM on a square lattice, showing asymmetric behaviour around half doping. While magnon dispersion is of Heisenberg form for (a) $n$=0.35, the anomalous magnon energy suppression at X relative to M for (b) $n$=0.7 indicates strong non-Heisenberg behavior, although the dispersion in $\Gamma$-X direction (c) is of nearly Heisenberg form.}
\label{dks5}
\end{figure}


\section{Results}

Figs. \ref{dks5} (a) and (b) show the renormalized magnon dispersion for a square lattice at two different band fillings, clearly showing asymmetric behavior about half doping. In the following, we have set $t=1$ as unit of energy scale. For $n$=0.35, magnon dispersions for different $U$ values almost completely overlap, in accordance with the result obtained above. Moreover, the dispersion is of the Heisenberg form. This band filling was not studied in the earlier investigations.\cite{golosov_2005,kapetanakis_2007} 

For $n$=0.7, the magnon dispersion does show an overall non-Heisenberg feature ($\omega_X < \omega_M/2$), with magnon energies at X anomalously suppressed in relation to M. The non-Heisenberg behavior weakens with increasing $U$ in agreement with the variational approach.\cite{kapetanakis_2007} However, the detailed nature of the dispersion is not in accordance with experiments. As seen in Fig. \ref{dks5} (c), comparison of magnon dispersion in $\Gamma$-X direction with a Heisenberg form having same spin stiffness shows only slight zone-boundary hardening, which is in contrast to the experimental findings of zone-boundary softening. 


Even for the uncorrelated FKLM, spin-charge coupling yields a non-Heisenberg magnon self energy (Eq. 18) for $n$$\gtrsim$0.5, resulting in anomalous suppression at X. The calculated finite-$U$ behavior follows from the weak suppression of magnon self energy with $U$, as explained below. In this filling regime, the distribution of $\chi^0 _\beta$ values is no longer sharply peaked, and a better approximation is therefore to replace $\chi^0 _\beta ({\bf q})$ by $\chi^0 _\beta (0)$ in Eq. (17), which yields:
\begin{eqnarray}
\Gamma({\bf k};{\bf q},\omega;{\bf Q},\Omega) &=& J \left (1-\frac {2JS + Um}
{\epsilon_{\bf k-q}^{\downarrow +} - \epsilon_{\bf k}^{\uparrow -} +\omega -i \eta} \right) \nonumber \\
& \approx & J \left ( \frac{\epsilon_{\bf k-q} - \epsilon_{\bf k} }{2JS + Um} \right ) 
\end{eqnarray}
showing the weak suppression of the spin-charge coupling strength and hence the magnon self energy with $U$.

\begin{figure}
\begin{center}
\vspace*{-2mm}
\hspace*{0mm}
\psfig{figure=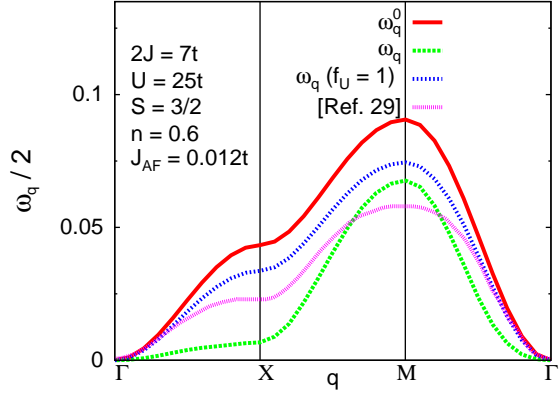,width=55mm,angle=-90}
\vspace*{-5mm}
\end{center}
\caption{Comparison of the calculated magnon dispersion with the three-body variational approach for the same set of parameters. Also shown is the result obtained by setting the factor $f_U$=1, highlighting the importance of correlation effects.}
\label{dks6}
\end{figure}  

Fig. \ref{dks6} shows comparison of the magnon self-energy approach with the three-body variational approach for the same set of parameters. The RPA level magnon dispersion shown here, obtained by including the scale factor $2S/(2S+n)$ and subtracting the contribution $ z J_{\rm AF}  S(1- \gamma_{\bf q}) $ due to the AF interaction between Mn spins, matches exactly with Ref. [29]. Although the RPA dispersion has nearly Heisenberg form, renormalized magnon energies are strongly suppressed at X. Compared to the variational approach, magnon energy suppression is significantly more at X and less at M, highlighting the importance of correlation effects in the low-$J$ regime. 

\begin{figure}
\begin{center}
\vspace*{-2mm}
\hspace*{0mm}
\psfig{figure=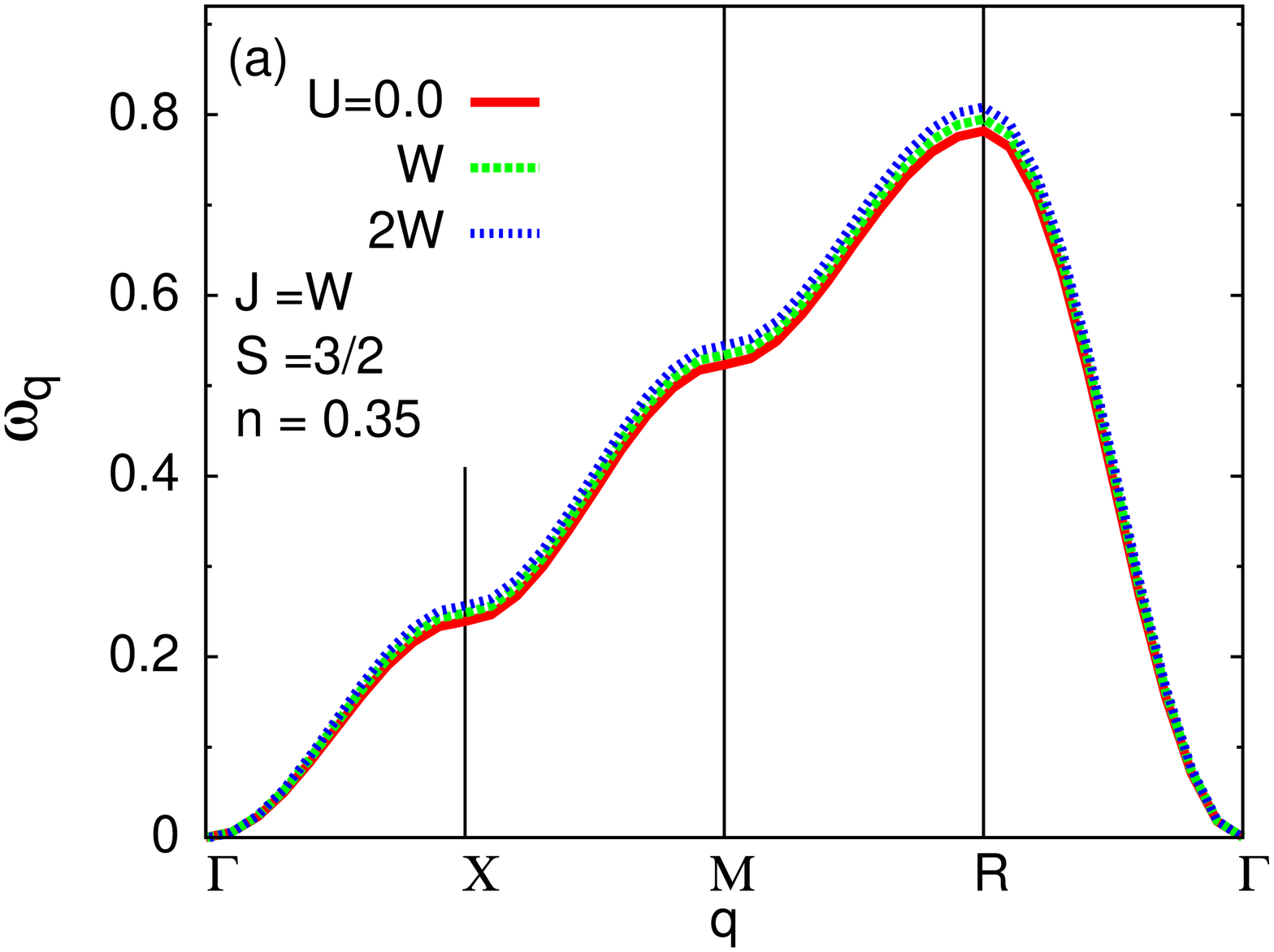,width=55mm,angle=-90}
\psfig{figure=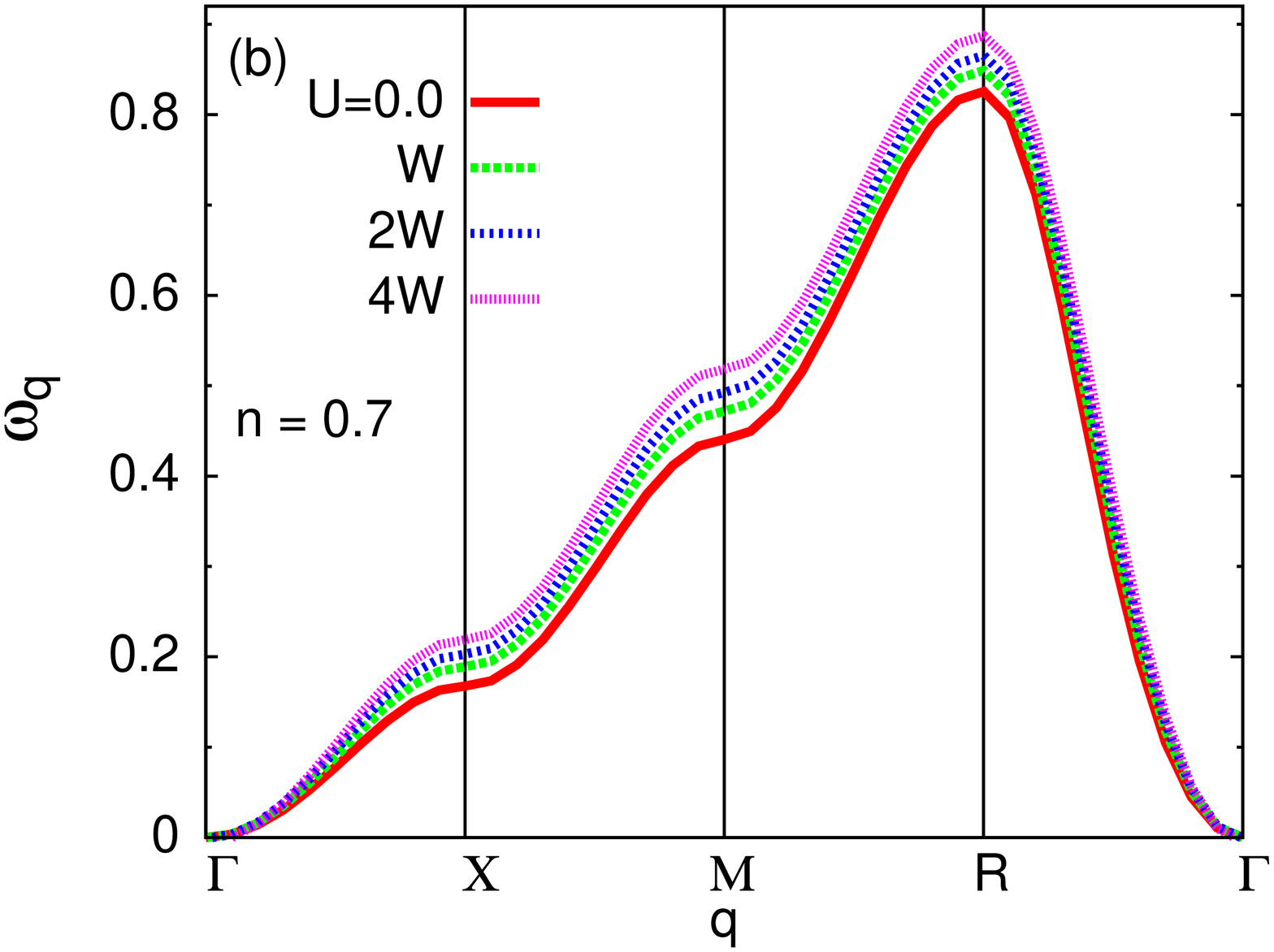,width=55mm,angle=-90}
\vspace*{-5mm}
\end{center}
\caption{Renormalized magnon dispersion for a two-orbital correlated FKLM on a cubic lattice for (a) $n$=0.35 per orbital, showing nearly Heisenberg form and (b) $n$=0.7 per orbital, showing significant deviation from Heisenberg form, though less pronounced than for the square lattice.}
\label{dks7}
\end{figure}  

The spin-charge coupling effect on magnon excitations can be readily extended to the two-band correlated FKLM involving spin-fermion interaction $- J \sum_i {\bf S}_{i}.({\mbox{\boldmath $\sigma$}}_{i\beta} + {\mbox{\boldmath $\sigma$}}_{i\gamma})$. As the two orbitals do not mix, both the bare magnon energy as well as the magnon self energy simply get doubled due to the two independent contributions to the effective spin couplings mediated by the particle-hole propagators involving the mobile $\beta$ and $\gamma$ electrons. Fig. \ref{dks7} shows renormalized magnon dispersion for the two orbital model in 3D, which was not investigated in earlier studies.\cite{golosov_2005,kapetanakis_2007} The results are qualitatively similar to the 2D cases. Near optimal filling $n$$\approx$0.35, corresponding to hole doping $x$=$1-2n$$\approx$0.3 for optimally doped manganites, the dispersion is of nearly Heisenberg form and almost independent of $U$.

\begin{figure}
\begin{center}
\vspace*{-2mm}
\hspace*{0mm}
\psfig{figure=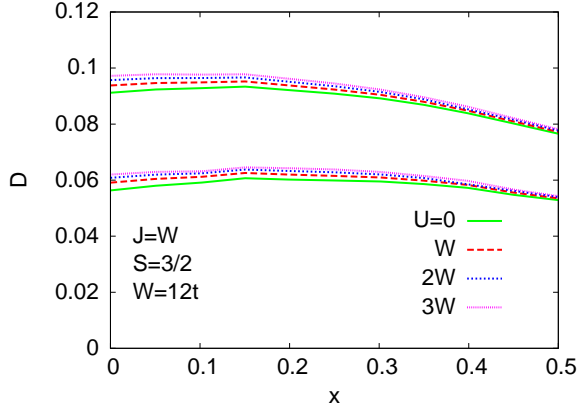,width=56mm,angle=-90}
\vspace*{-5mm}
\end{center}
\caption{Doping dependence of RPA (upper set) and renormalized (lower set) spin stiffness for different $U$ values, showing negligible dependence on $U$. The renormalized spin stiffness changes only slightly with hole doping, in agreement with neutron scattering results.}
\label{dks8}
\end{figure}

Fig. \ref{dks8} shows both RPA and renormalized spin stiffness as function of hole doping for the two-orbital model in 3d for different $U$ values. The corresponding band filling range is 0.25$\le$$n$$\le$0.5 per orbital. While quantum corrections beyond RPA level strongly renormalize the spin stiffness, it increases only slightly with $U$, as expected from Fig. \ref{dks7} (a). The spin stiffness exhibits a weak doping dependence, which is in agreement with the experimental findings. In contrast, in the one-band model, a strong doping dependence of spin stiffness near optimal band filling $n$$\approx$0.7 has been reported.\cite{kapetanakis_2007}

With reference to the experimentally observed anomalous zone-boundary magnon softening, earlier investigations have highlighted the role of spin-orbital coupling effects on magnon excitations in orbitally degenerate metallic ferromagnets.\cite{dheeraj1_2010}
While spin-charge coupling due to both finite $J$ and $U$ yield magnon damping \cite{dheeraj0_2010} and strong suppression of magnon energy in $\Gamma$-X direction, nevertheless this effect cannot account for the anomalous zone-boundary softening. It is
only on including inter-orbital interaction and a new class of spin-orbital coupling diagrams that low-energy staggered orbital fluctuations, particularly with momentum near $(\pi/2,\pi/2,0)$ corresponding to CE-type orbital correlations, is found to generically yield strong intrinsically non-Heisenberg $(1-\cos q)^2$ magnon self energy correction, resulting in strongly suppressed zone-boundary magnon energies in the $\rm\Gamma$-X direction \cite{dheeraj1_2010}.

\section{conclusions}

The role of Coulomb interaction on quantum corrections, spin-charge coupling effect, and magnon self energy in the correlated FKLM were investigated in terms of a purely fermionic representation which treated both Hund's coupling and Hubbard correlation on an equal footing, and allowed for a conventional many-body diagrammatic analysis. The systematic expansion scheme employed to incorporate correlation effects in the form of self-energy and vertex corrections explicitly preserved the continuous spin rotation symmetry and hence the Goldstone mode. Allowing for a continuous interpolation between the weak and strong coupling regimes, this approach is particularly suited for ferromagnetic manganites. 

The spin-charge coupling structure was extended for the correlated FKLM to include the additional correlation-induced self-energy and vertex correction diagrams for the magnon self energy. However, for the physically relevant intermediate coupling regime and optimal band filling ($n$$\approx$0.35) appropriate for optimally doped ($x$$\approx$0.3) manganites with doped holes shared between two degenerate $\rm e_g$ orbitals, due to a near cancellation of the correlation terms, the magnon self energy for the correlated FKLM is nearly same as for the uncorrelated FKLM, and the renormalized magnon energies nearly overlap for different $U$ values in both two and three dimensions. This is in contrast to several earlier investigations which emphasized the role of $U$ in the context of the several magnon anomalies observed in ferromagnetic manganites. In contrast, the present investigation of correlation induced spin-charge coupling effects, extended to three dimension and to the two-orbital model, does not show any zone-boundary magnon softening.

For the band filling $n$=0.7 case, which was considered in earlier investigations of the one-band model with $n$=$1-x$, the  renormalized magnon dispersion did show a pronounced non-Heisenberg feature, particularly in two dimensions ($\omega_X < \omega_M/2$), which weakened with increasing $U$. However, strictly in the $\Gamma$-X direction, the dispersion retained nearly Heisenberg form $(1-\cos q_x)$. Therefore, although the spin-charge coupling effect does yield magnon damping and anomalous magnon energy suppression at X in relation to M (or R in 3d) for $n$$\gtrsim$0.5, nevertheless this effect cannot account for the experimentally observed anomalous zone-boundary softening.

Hence, distinction between overall non-Heisenberg behavior and experimentally observed zone-boundary magnon softening is important. The latter requires additional $-(1- \cos q_x)^2$ term in the $\Gamma$-X direction, which leaves the spin stiffness unchanged and only lowers the magnon energy near the X point. The dominant non-Heisenberg behavior arises from contributions of the form $\cos q_x . \cos q_y$ etc. and $\cos q_x . \cos q_y . \cos q_z$ due to 2nd and 3rd neighbor effective spin couplings, which yield purely Heisenberg behavior in the $\Gamma$-X direction. 

Therefore, of the three major anomalies in magnon excitations in ferromagnetic manganites: (i) magnon damping results from the spin-charge coupling, (ii) almost constant spin stiffness with respect to hole doping is obtained in both spin-charge and spin-orbital coupling effects, and (iii) doping dependent zone-boundary softening must be ascribed to the spin-orbital coupling. 

\newpage
\section*{Appendix}

The RPA-level magnon propagators involving ladder sums are obtained as:
\begin{equation}
[\chi^{-+}_{\alpha\alpha}]_{\rm RPA}({\bf Q},\Omega) = \frac {\chi^0_\alpha(\Omega)(1-U\chi^0_\beta({\bf Q},\Omega))}{[1 - U_\alpha \chi^0_\alpha(\Omega)][1-U\chi^0_\beta({\bf Q},\Omega)] - J^2 \chi^0_\alpha(\Omega) \chi^0_\beta({\bf Q},\Omega)} 
\end{equation}
\begin{equation}
[\chi^{-+}_{\beta\beta}]_{\rm RPA}({\bf Q},\Omega) = \frac {\chi^0_\beta({\bf Q},\Omega)(1-U_\alpha \chi^0_\alpha(\Omega))}{[1 - U_\alpha \chi^0_\alpha(\Omega)][1-U\chi^0_\beta({\bf Q},\Omega)] - J^2 \chi^0_\alpha(\Omega) \chi^0_\beta({\bf Q},\Omega)} 
\end{equation}
\begin{equation}
[\chi^{-+}_{\beta\alpha}]_{\rm RPA}({\bf Q},\Omega) = \frac {J \chi^0_\beta({\bf Q},\Omega) \chi^0_\alpha(\Omega)}
{[1 - U_\alpha \chi^0_\alpha(\Omega)][1-U \chi^0_\beta({\bf Q},\Omega)] - J^2 \chi^0_\alpha(\Omega) \chi^0_\beta({\bf Q},\Omega)}
\end{equation}
from Eqs. (8-10) with $\phi_{\mu\mu'} = \chi^0_{\mu}\delta_{\mu\mu'}$ in Eq. (7). The corresponding RPA-level interaction vertices which appear in the quantum correction diagrams (Fig. 2) are similarly obtained as:
\begin{equation}
[\Gamma^{-+}_{\alpha\alpha}]_{\rm RPA}({\bf Q},\Omega) = [[\chi^{-+}_{\alpha\alpha}]_{\rm RPA}({\bf Q},\Omega)
-\chi^0_\alpha(\Omega)][\chi^0_\alpha(\Omega)]^{-2} 
\end{equation} 
\begin{equation}
[\Gamma^{-+}_{\beta\beta}]_{\rm RPA}({\bf Q},\Omega) = [[\chi^{-+}_{\beta\beta}]_{\rm RPA}({\bf Q},\Omega)
-\chi^0_\beta({\bf Q},\Omega)][\chi^0_\beta({\bf Q},\Omega)]^{-2} \nonumber\\
\end{equation}
\begin{equation}
[\Gamma^{-+}_{\beta\alpha}]_{\rm RPA}({\bf Q},\Omega) =  [\chi^{-+}_{\beta\alpha}]_{\rm RPA}({\bf Q},\Omega)
[\chi^0_\alpha(\Omega) \chi^0_\beta({\bf Q},\Omega)]^{-1} .
\end{equation}

These components involve simple relationships which are useful in compacting the nine quantum correction expressions (Eq. 12) into the single spin-charge coupling structure (Eqs. 18-21). From the magnon pole condition, it follows that 
\begin{equation}
J^2 \left ( \frac{\chi^0 _\alpha}{1-U_\alpha \chi^0 _\alpha} \right ) = \frac{1 - U\chi^0 _\beta}{\chi^0 _\beta} 
\end{equation}
substituting which in Eq. (35) yields:
\begin{equation}
\Gamma_{\beta\beta}^{-+} = J^2 \chi^{-+} _{\alpha \alpha} [1 - U\chi^0 _\beta]^{-2}  \; .
\end{equation}
Similarly, for the combination appearing in the (d)+(e)+(f) term in Eq. (12):
\begin{eqnarray}
J^2 \chi^{-+}_{\alpha\alpha} + 2JU \chi^{-+}_{\alpha\beta} + U^2 \chi^{-+}_{\beta\beta} &=& 
\frac{ J^2 \left ( \frac{\chi^0_\alpha}{1-U_\alpha \chi^0_\alpha} \right ) (1+U\chi^0_\beta) + U^2 \chi^0_\beta }
{ (1-U\chi^0_\beta) - J^2 \left ( \frac{\chi^0_\alpha}{1-U_\alpha \chi^0_\alpha} \right ) \chi^0_\beta } \; . \nonumber \\
&=& \Gamma_{\beta\beta}^{-+}
\end{eqnarray}
And finally for the [c] term in Eq. (12) involving $\Gamma^{-+}_{\beta\alpha}$:
\begin{eqnarray}
J \chi^0_\alpha \Gamma_{\beta\alpha}^{-+} &=& \frac{ J^2 \left ( \frac{\chi^0_\alpha}{1-U_\alpha \chi^0_\alpha} \right ) }
{ (1-U\chi^0_\beta) - J^2 \left ( \frac{\chi^0_\alpha}{1-U_\alpha \chi^0_\alpha} \right ) \chi^0_\beta } \; . \nonumber \\
&=& J^2 \chi^{-+}_{\alpha\alpha} [1-U\chi^0_\beta]^{-1}
\end{eqnarray}

Finite $U$ effects on the RPA level magnon energy (Eq. 15) are discussed below.
Since $[\chi^0 _\beta (0) - \chi^0 _\beta({\bf q})]$ is of order $10^{-4}$ for the parameters considered, Eq. (15) approximately reduces to:
\begin{eqnarray}
\omega_{\bf q} ^0 & \approx & J^2 (2S) [1-U\chi^0 _\beta(0)]^{-2} [ \chi^0 _\beta (0) - \chi^0 _\beta({\bf q}) ] \nonumber \\
& = & J^2 (2S) \left ( \frac{2JS + Um}{2JS} \right )^2 [ \chi^0 _\beta (0) - \chi^0 _\beta({\bf q}) ] \\
& \approx & J^2 (2S) [ \chi^0 _\beta (0) - \chi^0 _\beta({\bf q}) ]_{U=0} \; .
\end{eqnarray}
The last expression involves $\chi^0 _\beta$ difference for the uncorrelated FKLM with exchange gap $2JS$. Therefore, the RPA level magnon dispersion retains the Heisenberg form and energies as for the uncorrelated FKLM. This is explicitly shown below for the RPA level spin stiffness.  

Expanding the $\chi^0_\beta$ difference for small $q$ yields:
\begin{equation}
\chi^0_\beta(0) - \chi^0_\beta({\bf q}) = \left( \frac{1}{2JS + Um} \right )^2 \sum_{\bf k}  \left [ 
\frac{1}{2} ({\bf q}. {\mbox{\boldmath $\nabla$}} )^2 \epsilon_{\bf k} - \frac{({\bf q}.{\mbox{\boldmath $\nabla$}} \epsilon_{\bf k})^2}{2JS + Um} \right ]  
\end{equation}
from which the spin stiffness in $d$ dimensions is obtained as:
\begin{equation}
D^{(0)} = \omega_{\bf q} ^{(0)} / q^2 = \frac{1}{2S} \frac{1}{d} \sum_{\bf k} \left [\frac{1}{2} 
{\mbox{\boldmath $\nabla$}}^2 \epsilon_{\bf k}  - 
\frac{ ({\mbox{\boldmath $\nabla$}} \epsilon_{\bf k})^2 }{2JS +Um} \right ] \; .
\end{equation}
The first term (delocalization energy loss upon spin twisting) is the dominant contribution at low band filling $(m \ll 1)$,
and is independent of $U$, whereas the second term (exchange energy gain) results in a weak enhancement of the spin stiffness with $U$.

\end{document}